\documentclass[aps,
        twocolumn,
        superscriptaddress,
        10pt]{revtex4-1}

\usepackage{amsmath}
\usepackage{graphicx}
\usepackage{bm}
\usepackage{gensymb,textcomp}
\usepackage{times}
\usepackage{hyperref}
\usepackage{color}
\hypersetup{pdfpagemode=UseNone}
\usepackage{mathrsfs}

\newcommand{\nhat}{\hat{n}}
\newcommand{\xhat}{\hat{x}}
\newcommand{\yhat}{\hat{y}}
\newcommand{\zhat}{\hat{z}}

\newcommand{\what}{\hat{w}}

\newcommand{\bE}{\mathbf{E}}
\newcommand{\bH}{\mathbf{H}}
\newcommand{\bJ}{\mathbf{J}}

\newcommand{\bX}{\mathbf{X}}

\newcommand{\bZ}{\mathbf{Z}}
\newcommand{\TE}{\text{TE}}
\newcommand{\TM}{\text{TM}}
\renewcommand{\t}[1]{\text{#1}}

\renewcommand{\Re}{\operatorname{Re}}
\renewcommand{\Im}{\operatorname{Im}}

\begin{document}

\title{Arbitrary beam control using passive lossless metasurfaces
enabled by orthogonally-polarized custom surface waves}

\author{Do-Hoon Kwon}
\email{dhkwon@umass.edu}
\affiliation{Department of Electrical and Computer Engineering,
University of Massachusetts Amherst, Amherst, Massachusetts 01003, USA}
\author{Sergei A. Tretyakov}
\affiliation{Department of Electronics and Nanoengineering,
Aalto University, P.O. Box 15500, 00076 Aalto, Finland}

\begin{abstract}
For passive, lossless
impenetrable metasurfaces, a design technique for arbitrary
beam control of receiving, guiding, and launching is presented.
Arbitrary control is enabled by a custom surface
wave in an orthogonal polarization
such that its addition to
the incident (input) and the desired 
scattered (output) fields is supported by a reactive surface
impedance everywhere on the reflecting surface. Such a custom
surface wave (SW) takes the form of an evanescent wave propagating
along the surface with a spatially varying envelope.
A growing SW appears when an illuminating beam
is received. The SW amplitude stays constant
when power is guided along the surface. The amplitude diminishes
as a propagating wave (PW) is launched from the surface as a
leaky wave.
The resulting reactive tensor impedance profile
may be realized as an array of anisotropic metallic resonators
printed on a grounded dielectric substrate. Illustrative design
examples of a Gaussian beam translator-reflector, a
probe-fed beam launcher, and a near-field focusing lens are provided.
\end{abstract}

\maketitle

\section{Introduction}
\label{sec:introduction}

Comprising an electrically thin layer of subwavelength
resonators~\cite{holloway_ieeemap2012}, 
metasurfaces are capable of tailoring the
characteristics of propagating and evanescent waves
with a significantly reduced loss compared with traditional
volumetric metamaterials. In particular, gradient metasurfaces,
those with spatially varying surface parameters, have received
increased attention in recent years. By imparting
a linearly gradient discontinuity on a reflection
or transmission phase,
the incident wave can be deflected anomalously upon reflection
or transmission. Since a recent formalization of the generalized laws
of reflection and refraction~\cite{yu_science2011}, a wide
variety of novel metasurface applications have been proposed
and demonstrated.
Using penetrable metasurfaces, anomalous refraction has
been demonstrated from microwave to optical
regimes~\cite{yu_science2011,aieta_nanolett2012b,
pfeiffer_prl2013,wong_ieeejmtt2015,asadchy_prl2015}. Introducing
a proper nonlinear phase distribution leads to
lenses~\cite{monticone_prl2013,lin_science2014,wang_advoptmater2015,
khorasaninejad_science2016}. Holograms are designed by
locally controlling the transmission amplitude and/or
phase~\cite{ni_natcommun2013,huang_natcommun2013}. A careful
geometrical arrangement of transmission blocks 
having different transmission phases can
create optical vortex beams~\cite{yu_science2011,
genevet_apl2012,karimi_lsa2014,shalaev_nanolett2015,
chong_nanolett2015}.

A variety of novel functionalities have also been demonstrated
in the reflection mode or for impenetrable metasurfaces.
Anomalous reflection described by the generalized Snell's law
has been in use for reflectarray designs in the antenna engineering
community~\cite{huang2008}.
Reflection-mode lenses~\cite{pors_nanolett2013,asadchy_prl2015},
holograms~\cite{zheng_natnanotech2015,wen_natcommun2015}, and
optical vortex generator~\cite{yang_nanolett2014,yu_apl2016}
have been reported. In \cite{sun_natmater2012}, conversion
from a plane wave into an SW was cast as a special
case of anomalous reflection, wherein the reflected wave vector
enters the evanescent (invisible) spectral range. This work
has spurred a strong interest in the conversion process and
efficiency between propagating and evanescent 
waves~\cite{sun_nanolett2012,wang_apl2012,pors_lsa2014,fan_ieeejap2016},
either
in the form of surface plasmon polariton waves at optical
frequencies or SWs at microwave frequencies.
In addition, gradient metasurfaces raised
a renewed interest in improving traditional leaky-wave
antennas based on arrays of subwavelength printed resonators
on a grounded dielectric
substrate~\cite{fong_ieeejap2010,patel_ieeejap2011,
minatti_ieeejap2016,minatti_ieeejap2016b}.
{\color{red}For nonlinear metasurfaces, design of proper 
phase distributions associated with a judicious choice of
nonlinear meta-atom arrangement can manipulate beams such as
steering, splitting, focusing, and vortex generation at harmonic
frequencies~\cite{segal_natphoton2015,wolf_natcommun2015,
almeida_natcommun2016,tymchenko_prb2016}. 
If the required grating period allows no more than 
three propagating Floquet modes, anomalous reflection 
with nearly unitary efficiency can be realized using 
a few or even a single inclusion per period, 
using blazed gratings \cite{loewen_applopt1977}
or recent conceptualizations such as
meta-gratings \cite{radi_prl2017} 
and agressively discretized metasurfaces 
\cite{wong_arxiv2017}.} 

Along with the designs and demonstrations of novel wave transformations,
their efficiencies started to receive attention.
Toward complete wavefront tailoring via a complete $2\pi$
transmission phase coverage, the generalized Snell's laws were
demonstrated using cross-polarized light in
\cite{yu_science2011,aieta_nanolett2012b}. It was theoretically
revealed that the maximum power coupling efficiency is
only 25\%~\cite{monticone_prl2013}. Utilizing a balanced
pair of induced tangential electric and magnetic dipole moments,
Huygens' metasurfaces~\cite{pfeiffer_prl2013,wong_ieeejmtt2015}
can achieve polarization-preserving full transmission with
an arbitrary phase in a complete $2\pi$ span. 
{\color{red}At microwave frequencies, low-loss conductor
trances can be arranged to realize the electric and magnetic
dipoles. In the optical regime, a careful choice of the dimensions,
shape, material, and periodicity for an array of
dielectric resonator meta-atoms
can bring the electric and magnetic dipole resonances 
together~\cite{decker_advoptmat2015,campione_opex2015,liu_nanolett2017,
arslan_jphysd2017}.}
However,
requiring perfect anomalous transmission without reflection
using Huygens' metasurfaces results in globally lossless, but
locally active or lossy,
surface characterizations~\cite{epstein_josab2016}.
It was found that an $\Omega$-type bianisotropic metasurface
is capable of achieving perfect anomalous refraction without loss or
reflection~\cite{wong_ieeejawpl2016,epstein_ieeejap2016,
asadchy_prb2016}. Similarly for reflective metasurfaces,
an anomalous reflector based on the generalized Snell's
law~\cite{yu_science2011} cannot perfectly reflect an incident
plane wave into an anomalous direction, but necessarily
entails parasitic reflections into undesired 
directions~\cite{estakhri_prx2016,asadchy_prb2016}.
Again, requiring that the reflected wave
be a single plane wave in the desired anomalous direction
results in a strongly dispersive reflecting surface with local
active-lossy characteristics. Towards achieving perfect
reflection using passive metasurfaces, a reflector design approach
based on the leaky-wave antenna principle has been recently
reported~\cite{diaz-rubio_sciadv2017}, where 
a measured reflection power efficiency of 94\% has been realized
using a printed patch array implementation.

For penetrable metasurfaces, a design recipe for passive, lossless
$\Omega$-type bianisotropic metasurfaces has been 
introduced~\cite{epstein_ieeejap2016,epstein_prl2016}. It was
shown that an envisioned wave transformation is supported
by a lossless $\Omega$-type bianisotropic metasurface if power is
locally conserved. By introducing auxiliary SWs
for equalizing the power on both sides of the surface,
designs for a perfect beam split as well as anomalous
reflection and refraction have been numerically demonstrated.
The same design methodology was applied to a metasurface over
a perfect electric conductor surface for perfect
anomalous reflection~\cite{epstein_iwat2017}.
The design philosophy has been recently extended
to lossless, impenetrable surfaces characterized by a 
reactive surface impedance for perfect reflection control
of plane waves~\cite{kwon_prb2017},
where SWs of an orthogonal polarization were chosen.
Recently, a metasurface design study toward perfect conversion
between a plane wave and an SW has been 
reported~\cite{tcvetkova_arxiv2017}.
In \cite{achouri_metamaterials2016,achouri_arxiv2016},
metasurface designs for SW routing of beams were presented
for transmissive metasurfaces. Starting from the generalized
sheet transition conditions, spatially-dispersive susceptibilities
of the metasurface were specified in terms of the desired
field discontinuities. However, such metasurface designs
have spatially varying surface parameters with alternately
active and lossy properties, posing a challenge for realization.

This paper presents a design technique for passive, lossless
impenetrable metasurfaces for perfect beam manipulation.
It relies on synthesis of an evanescent auxiliary wave
such that the total fields on the metasurface have no real power
component through the surface everywhere. It is shown that
evanescent waves having a spatially growing and diminishing
envelope along the propagation direction can receive and launch PWs,
respectively.
The exact power profile of this custom SW is
determined for a given desired set of illuminating (input) and
radiating (output) waves via an efficient numerical 
optimization procedure. As a natural extension
of infinite, periodic metasurface designs for plane 
waves~\cite{kwon_prb2017},
a treatment of beam manipulation not only provides a design recipe for
transforming arbitrary incident waves having a continuum of
spectrum, but also demonstrates the versatility of metasurfaces
in wave manipulation that is achievable with passive realizations.
The design produces an inhomogeneous, anisotropic
surface reactance tensor,
which may be approximately realized using the standard printed circuit
technologies.

\section{Beam control: perfect receiving, guiding, and launching}
\label{sec:beam_control}

\begin{figure}[t]
 \centering
 \includegraphics*{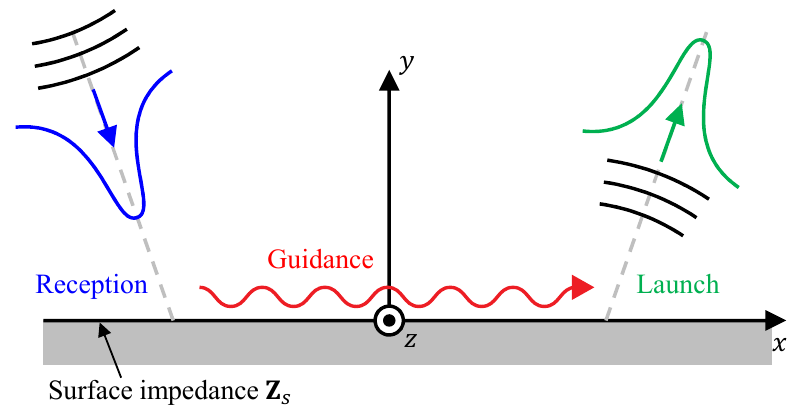}
 \caption{Reception, guidance, and launch of an incident beam
by a planar metasurface.
{\color{red}The metasurface at $y=0$ is impenetrable and characterized
by a position-dependent surface impedance $\bZ_s$.}}
 \label{fig:beam_control}
\end{figure}

Figure~\ref{fig:beam_control} illustrates the design problem
for an impenetrable metasurface under consideration. 
In free space, a known beam wave illuminates a planar 
impenetrable surface in the $xz$-plane. 
Using a passive, lossless metasurface, it desired that
the incident beam is
first received and converted into an SW,
subsequently guided along the surface, before it is
launched back into space from a location that is electrically
far separated from the receiving point.
The power of the desired output wave
is the same as that of the incident beam. Furthermore,
a surface with reciprocal parameters is desired.

At the heart of the design lies perfect conversion of
a PW into an SW and a subsequent
transition into lossless guiding along the surface.
Launching is a reciprocal process of receiving and a
planar structure with a uniform reactive surface impedance can
guide an SW without loss. For such a converter
based on the generalized Snell's law, it was found that
an SW, once generated, cannot propagate along
the gradient metasurface without suffering secondary
scattering by the same metasurface back into space.
This lowers the conversion efficiency for electrically long
converters~\cite{qu_epl2013}, defined for the power accepted by
a homogeneous guiding metasurface. The conversion efficiency
is significantly improved by introducing a meta-coupler at a height
over a homogeneous surface that supports a surface 
wave~\cite{sun_lsa2016}. A drawback of this approach is
an increased overall thickness. Nevertheless, the conversion efficiency
cannot reach 100\% because some of the generated SW
fields still undergo secondary scattering by the meta-coupler.

Instead of relying on the generalized reflection law,
designing the exact SW tailored to the
incident wave can lead to a metasurface design that is capable
of perfect PW-to-SW conversion.
With respect to the chosen guided direction
(the $+x$-axis direction in Fig.~\ref{fig:beam_control}), 
this custom SW must have a position-dependent amplitude profile
such that the amplitude grows over the range of beam illumination
due to PW-to-SW conversion, stays constant over the guided range
without leakage radiation, then diminishes over the range of
beam launch due to SW-to-PW conversion. Over the beam receiving
range, the PW is continuously converted into the SW, so 
there is no abrupt junction between conversion and guiding ranges.
The exact profile of the SW is determined such that the
PW-to-SW conversion is done in a locally
lossless manner, without loss in
or power supplied by the metasurface, everywhere on the surface.
Once the exact SW is designed, the boundary condition for the
impenetrable surface is found for supporting the envisioned
total fields as a reactive surface impedance.

In this work, the same polarization is chosen between the
input and output beams. For the surface wave, an orthogonal
polarization is chosen as in \cite{kwon_prb2017,tcvetkova_arxiv2017}.
An orthogonally polarized SW avoids creating an interference
power pattern between the PW and SW on the surface,
simplifying the SW synthesis significantly.
In the following sections, beam controlling
metasurfaces in two-dimensional (2-D) configurations 
with PWs in the
TE polarization (with respect to $y$)
and SWs in the TM polarization are presented. 

\section{Impenetrable surface characterization}
\label{sec:surf_imped}

Let us adopt and suppress an $e^{j\omega t}$ time convention 
for the following time-harmonic analysis at an angular frequency 
$\omega$.
Referring to Fig.~\ref{fig:beam_control}, the total fields
$\bE=\xhat E_x+\yhat E_y+\zhat E_z$, $\bH=\xhat H_x+\yhat H_y+\zhat H_z$
in $y\geq 0$
comprise a superposition of those from the incident beam,
the SW bound to the metasurface at $y=0$, and the output beam.
For an impenetrable surface characterized by a surface impedance
$\bZ_s$, the tangential electric and magnetic fields on the surface,
$\bE_t=\xhat E_{tx}+\zhat E_{tz}$ and $\bH_t=\xhat H_{tx}+\zhat H_{tz}$,
are related by
\begin{equation}
\bE_t=\bZ_s\bJ_s=\bZ_s\nhat\times\bH_t,
\label{ibc}
\end{equation}
where $\nhat=\yhat$ is the unit surface normal and
$\bJ_s=\nhat\times\bH_t$ is the induced electric surface current
density at $y=0$. The rank-2 tensor $\bZ_s$ may be written
as a 2\texttimes2 matrix in the $xy$-plane.

Toward arriving at a lossless and reciprocal surface specification,
consider $\bZ_s$ in terms of a reactance tensor $\bX_s$
written by
\begin{equation}
\bZ_s=j\bX_s=j
\begin{bmatrix}
X_{xx} & X_{xz}\\
X_{zx} & X_{zz}
\end{bmatrix},
\label{def:zs_as_jxs}
\end{equation}
where the four reactance elements are real. Using (\ref{def:zs_as_jxs})
in (\ref{ibc}), matching both the real and imaginary parts on the two
sides uniquely determines the four tensor elements
as~\cite{kwon_prb2017}
\begin{equation}
\bX_s=\frac{1}{\Im\{H_{tx}H_{tz}^*\}}
\begin{bmatrix}
\Re\{E_{tx}H_{tx}^*\} & \Re\{E_{tx}H_{tz}^*\}\\
\Re\{E_{tz}H_{tx}^*\} & \Re\{E_{tz}H_{tz}^*\}
\end{bmatrix}.
\label{xs:from_fields}
\end{equation}
This reactance tensor represents a lossless, reciprocal surface
if the off-diagonal terms are equal to each other. This condition
is equivalent to~\cite{kwon_prb2017}
\begin{equation}
S_n=\yhat\cdot\frac{1}{2}\Re\{\bE_t\times\bH_t^*\}=0,
\label{eq:Sn_zero}
\end{equation}
where $S_n$ is the normal component of the Poynting vector on the
surface. Hence, if there is no net power penetrating the surface
at a given point either as absorption by the surface $(S_n<0)$
or as power injected into $y>0$ by a source on the surface$(S_n>0)$, 
the surface is locally lossless and reciprocal. If (\ref{eq:Sn_zero})
is satisfied everywhere, the metasurface is lossless and reciprocal
globally.

The reactance tensor of a lossless surface is in general a
Hermitian tensor. Therefore, a real, symmetric tensor in
(\ref{xs:from_fields}) under the condition (\ref{eq:Sn_zero})
is not the unique solution for a lossless, reciprocal surface
that supports $(\bE,\bH)$
because the off-diagonal elements were set to have no
imaginary parts by design.
Still, a symmetric $\bX_s$ with real-valued
elements is advantageous because its eigenvalues are real and
the eigenvectors are orthogonal, allowing realization using
an array of rotated resonant meta-atoms.

\section{Synthesis of the surface wave}
\label{sec:sw_synthesis}

In Fig.~\ref{fig:beam_control}, consider a 2-D configuration
where the input and output PW fields are TE-polarized and the
SW field is TM-polarized with all the fields invariant
along the $z$-axis. Then, $E_z(x,y)$ and $H_z(x,y)$ uniquely
determine all remaining field components in the $xy$-plane
for the TE and TM polarizations, respectively.
The electromagnetic
uniqueness theorem allows the fields in the half space $y\geq 0$, 
which is bounded
by the metasurface and subject to the radiation condition at infinity,
to be uniquely determined from those on the metasurface at $y=0$.
Furthermore, the planar metasurface geometry makes analysis and
design in the spectral domain efficient via Fourier transform.
Derivations for field components and the normal component of the
Poynting vector are available in
Appendices~\ref{appendix:te} and \ref{appendix:tm} for TE-
and TM-mode fields, respectively.

\begin{figure}[t]
 \centering
 \includegraphics*{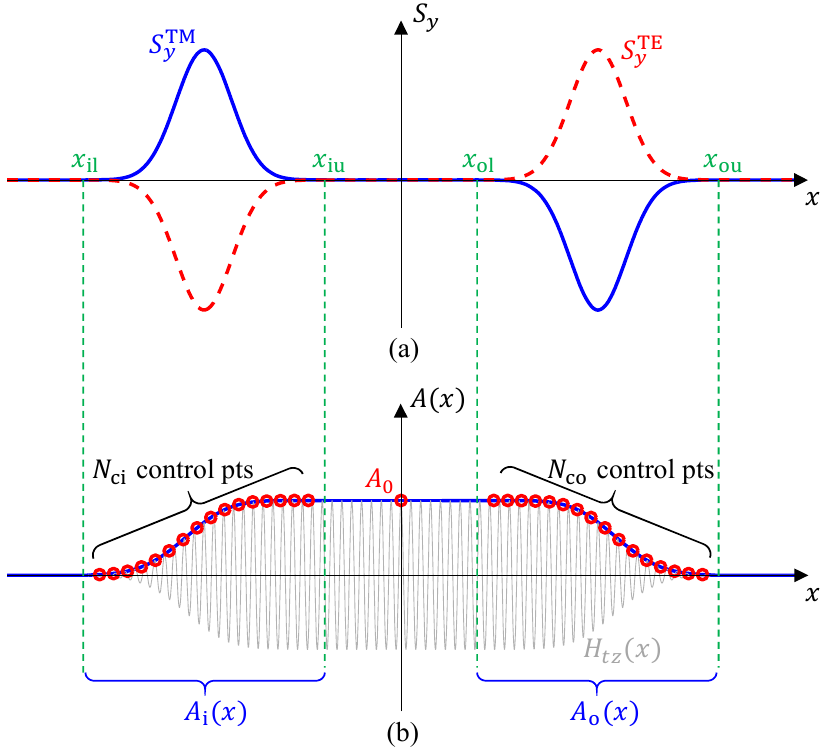}
 \caption{Illustration of
power and envelope profiles for SW synthesis.
(a) The normal component of the Poynting vector for the TE-
and TM-polarized fields.
(b) The envelope design for the TM-polarized SW component,
$H_{tz}(x)$.}
 \label{fig:StetmAenv}
\end{figure}

The metasurface design starts from a complete knowledge of the
incident beam and the desired output beam. 
For a given input beam, we assume that
the output PW fields are known from a separate overall function design
of the metasurface. In particular, 
the total output power should be equal to the input power in order
to make the overall system lossless. In terms of
$E_{tz}(x)$ and its Fourier transform 
$\tilde E_{tz}(k_x)=\mathscr{F}\{E_{tz}(x)\}$,
the expression for 
the normal component of the time-average Poynting vector
is given in (\ref{STEy}), which can be written compactly as
\begin{equation}
S^\TE_y(x)=\frac{1}{2k\eta}\Re\left\{E_{tz}^*(x)
\mathscr{F}^{-1}\left[k_y\tilde E_{tz}(k_x)\right]\right\}.
\label{STEy:compact}
\end{equation}
Typically, closed-form field expressions are approximate and
available only in the paraxial region near the beam axis, even for
commonly studied beams such as Gaussian or Bessel beams. Hence,
the primary quantities for analysis and design are $z$-directed
field components on the metasurface in this study.

For the TE-polarized input and output fields in a typical
beam control application described in Fig.~\ref{fig:beam_control},
the normal power density profile is illustrated in
Fig.~\ref{fig:StetmAenv}(a) as a red dashed curve. For a
beam wave having a finite support, we can define the lower and
upper $x$-limits for the input wave, $x_\t{il}$ and $x_\t{iu}$,
such that non-negligible fields and power are observed in
$x\in[x_\t{il},x_\t{iu}]$. Similarly, two $x$-coordinates
are set to define the range where the output beam leaves the
surface to be $x\in[x_\t{ol},x_\t{ou}]$. Over the illuminated
range, $S^\TE_y(x)<0$. Over the beam launch range, $S^\TE_y(x)>0$.

The objective of the SW synthesis is to design a custom
SW, specified by $H_{tz}(x)$ in $-\infty<x<\infty$,
such that (\ref{eq:Sn_zero}) is satisfied for all $x$.
Choosing an orthogonal TM polarization for the SW makes the
normal Poynting vector component of the total fields
equal to an algebraic sum of
the TE and TM polarization components, i.e.,
\begin{equation}
S_n(x)=S^\TE_y(x)+S^\TM_y(x).
\label{Sn_as_sum}
\end{equation}
The expression for the TM-mode Poynting vector component, 
shown in (\ref{STMy}), can be written concisely as
\begin{equation}
S^\TM_y(x)=\frac{\eta}{2k}\Re\left\{H_{tz}^*(x)
\mathscr{F}^{-1}\left[k_y\tilde H_{tz}(k_x)\right]\right\}.
\label{STMy:compact}
\end{equation}

In Fig.~\ref{fig:StetmAenv}(a), the target power profile for the
TM-polarized SW is shown as a solid blue curve. For $H_{tz}(x)$,
we choose propagating function in the $+x$ direction with a
custom $x$-dependent envelope. This envelope function is a
real function, taking non-negative values. It is denoted
by $A(x)$ and shown as a blue curve
in Fig.~\ref{fig:StetmAenv}(b).
Spatially, no SW is present to the left of the illumination range
in $x<x_\t{il}$ and past the launch range 
of the output PW in $x>x_\t{ou}$. Between complete reception of
the incident beam and initiation of the output beam launch,
in $x\in(x_\t{iu},x_\t{ol})$, the
SW propagates in the $+x$-direction, carrying
the received power without attenuation or growth.
During reception in
$x\in[x_\t{il},x_\t{iu}]$, the envelope monotonically increases
as power is gradually accumulated for the SW as the input PW is
converted into the SW. The reverse process occurs in the launch
range $x\in[x_\t{ol},x_\t{ou}]$. Hence, we express the envelope
function as
\begin{equation}
A(x)=\begin{cases}
0, & x<x_\t{il}\ \  \text{or}\ \  x>x_\t{ou}\\
A_\t{i}(x), & x_\t{il}\leq x\leq x_\t{iu}\\
A_0, & x_\t{iu}<x<x_\t{ol}\\
A_\t{o}(x), & x_\t{ol}\leq x\leq x_\t{ou}
\end{cases},
\end{equation}
where $A_\t{i}(x)$ is an increasing function associated with
the reception, $A_\t{o}(x)$ is a decreasing function associated with
the launch, and $A_0$ is a constant. Once $A(x)$ is specified,
the tangential magnetic field component on the surface
is defined as
\begin{equation}
H_{tz}(x)=A(x)e^{-jk_\t{c}x},
\label{def:Htz}
\end{equation}
where $k_\t{c}$ is the center (or ``carrier'') wavenumber of the SW
chosen in the invisible range $(k_\t{c}>k)$ so that the synthesized
TM wave propagates while bound to the metasurface. A snapshot
of $H_{tz}(x)$ (the real part) is visualized as a thin gray curve in
Fig.~\ref{fig:StetmAenv}(b).
It is stressed that $H_{tz}(x)$ in (\ref{def:Htz})
is defined on the surface $(y=0)$ only. In order to obtain 
the associated fields that are valid (i.e., Maxwellian)
in the entire volume of existence $(y\geq 0)$, we express
$H_{tz}(x)$ as a superposition of homogeneous and inhomogeneous
plane wave components evaluated at $y=0$, using Fourier expansions.

{\color{red}At this point,
it is instructive to assess the qualitative behavior
of the synthesized SW in $y\geq 0$. The definition of $H_{tz}(x)$
with respect to the envelope $A(x)$ in (\ref{def:Htz}) relates their
Fourier transforms as
\begin{equation}
\tilde H_{tz}(k_x)=\tilde A(k_x-k_c).
\label{def:tHtz}
\end{equation}
Since $A(x)$ is a real, non-negative function of $x$, $\tilde A(k_x)$
has the maximum at $k_x=0$. While a finite spatial support of $A(x)$
makes $\tilde A(k_x)$ not completely vanish at large $|k_x|$,
the effective spectral width of $\tilde A(k_x)$ is inversely
proportional to the range of non-zero $A(x)$.
From (\ref{Hz:invF}), the resulting H-field component of the SW
in $y\geq 0$ is expressed as
\begin{equation}
H_z(x,y)=\frac{1}{2\pi}\int\limits_{-\infty}^\infty
\tilde A(k_x-k_c)e^{-j(k_xx+k_yy)}dk_x.
\label{def:Hz}
\end{equation}
It can be seen that the dominant contribution to $H_z(x,y)$ comes
from its spectrum at and around $k_x=k_c$. At $k_x=k_c$,
the integrand in (\ref{def:Hz}) represents a
surface wave bound to the $y=0$ surface 
with a positive propagation constant of $k_x=k_c$ 
in the $x$-direction
and an attenuation constant of $|k_y|=\sqrt{k_c^2-k^2}$ in the
$y$-direction. Over the beam reception and launch ranges, the
amplitude of the SW grows and diminishes gradually with respect to $x$, 
respectively. In both ranges,
the SW propagates in the $+x$-direction and attenuates exponentially
in the $+y$-direction. Finally, contributions from all other spectral
ranges make the SW exist over the intended, finite range of $x$.}

Requiring $A(x)$ to be a continuous function, it remains to
determine the functions, $A_\t{i}(x)$ and $A_\t{o}(x)$, and
the constant $A_0$. For $A_\t{i}(x)$, we choose
a closely spaced set of $N_\t{ci}$ control
points of values $A_n$ at locations
$x=x_{\t{c}n}$ $(n=-N_\t{ci},-N_\t{ci}+1,\ldots,-1)$.
Similarly, $N_\t{co}$ number of control points of values
$A_n$ at $x=x_{\t{c}n}$ $(n=1,2,\ldots,N_\t{co})$ are defined
for $A_\t{o}(x)$.
The constant level at $A_0$ should make smooth connections to
the transition ranges on both sides.
A continuous envelope function is then defined using 
these control points (locations and values) via interpolation.
No control points are
assigned at the boundary locations of the input and output
beam ranges. 
In Fig.~\ref{fig:StetmAenv}(b), control points are indicated by
red circles. The number and $x$-locations
of the control points are set appropriately by the designer
depending on the allowed complexity of the synthesized SW.

Efficient numerical optimizations can determine the values of
$A_n$ $(n=-N_\t{ci},\ldots,N_\t{co})$. In this work, a square
error function defined by
\begin{equation}
\text{error}=\int\limits_{-\infty}^\infty
|S^\TE_y(x)+S^\TM_y(x)|^2dx
\label{def:sqerr}
\end{equation}
is minimized with respect to the $N_\t{ci}+N_\t{co}+1$ number
of control point values. Specifically, a gradient-free optimization
method available through the \textit{fminsearch} function in Matlab
has been used for all designs presented in Sec.~\ref{sec:designs}.
After some adjustments to the number and locations of control points,
all optimizations converged within a convergence tolerance of
$10^{-6}$ relative to the error associated with a null initial guess.

Once the optimized envelope function $A(x)$ is determined,
using $H_{tz}(x)$
from (\ref{def:Htz}) in (\ref{Ex:invF}) at $y=0$ finds
$E_{tx}(x)$. Together with $E_{tz}(x)$ and $H_{tx}(x)$
given by the TE-mode fields, they determine the surface
reactance tensor that achieves the designed beam control
via (\ref{xs:from_fields}).

Since a spatially varying envelope function is used, the synthesized
SW has a continuum of spectrum concentrated around $k_x=k_\t{c}$ rather
than a discrete spectrum. As a result, some of the spectrum will
spill into the visible region $(|k_x|<k)$. This means that
satisfaction of (\ref{eq:Sn_zero}) for all $x$, 
or equivalently the resulting
passive, lossless metasurface, necessarily entails excitation of
extra PW components in order to
to perform the required beam manipulation.
The amount of TM-polarized PW components can be assessed by
evaluating the total TM-mode power per unit length in the $z$-axis
direction that escapes the metasurface.
This power, denoted $P^\TM_y$, is found to have a spectral
integral representation given by
\begin{equation}
P^\TM_y=\int\limits_{-\infty}^\infty S^\TM_y(x)dx
=\int\limits_{-\infty}^\infty
\Re\left\{\frac{\eta k_y}{4\pi k}|\tilde H_{tz}(k_x)|^2\right\}dk_x.
\label{PTMy}
\end{equation}
It can be seen that non-zero TM-mode spectrum in the visible
region leads to some TM-mode power that is not bound to the
metasurface.
In principle, this amount of TM-polarized PW components can be
reduced to approach zero by increasing $k_\t{c}$ deep into the invisible
range. However, this
results in faster spatial variations for the surface impedance
tensor elements, making an accurate realization more challenging.

\section{Design examples}
\label{sec:designs}

{\color{red}In the following examples, a metasurface design is
characterized as a tensor surface impedance $\bZ_s(x)$, given
in terms of the reactance tensor elements in (\ref{def:zs_as_jxs}).
In numerical validations using COMSOL Multiphysics,
the impedance boundary condition
(\ref{ibc}) is enforced at $y=0$.}

\subsection{Gaussian beam translator-reflector}
\label{subsec:reflector}

\begin{figure}[t]
 \centering
 \includegraphics*{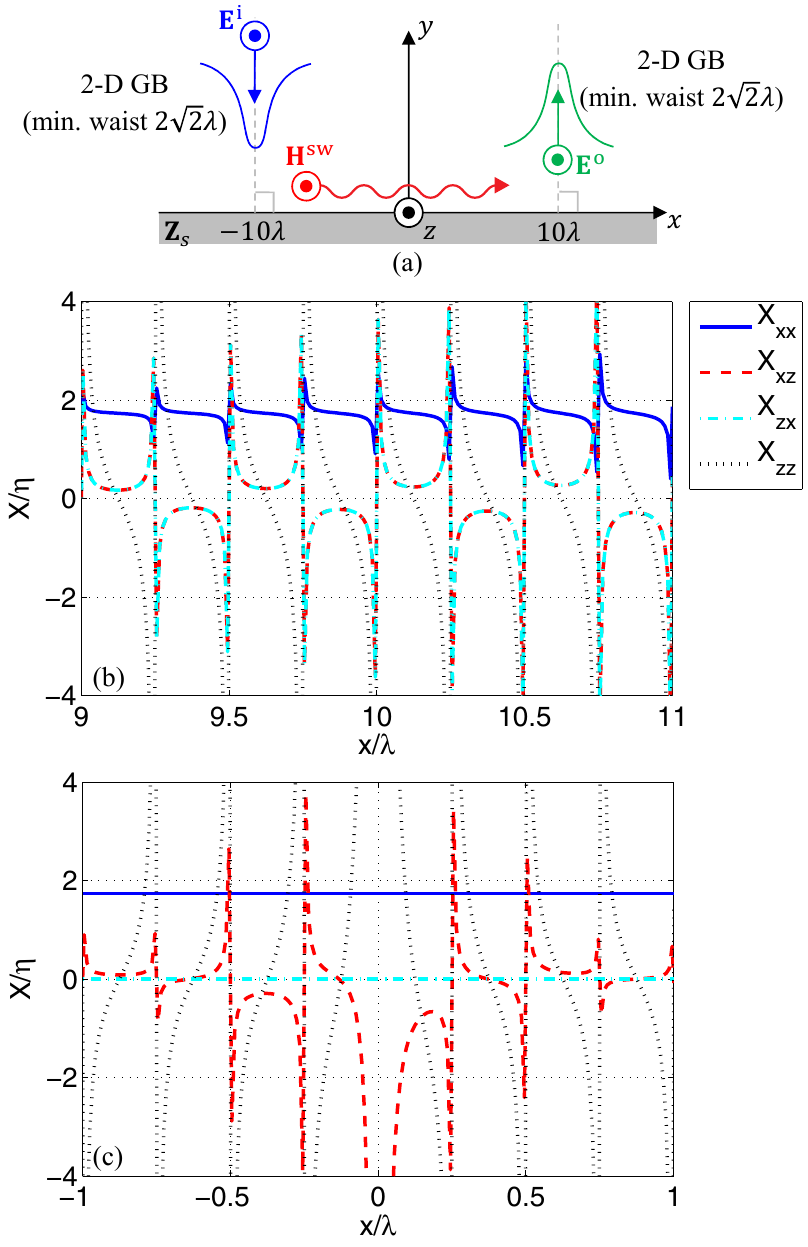}
 \caption{A GB translator-reflector design. 
(a) The design requirement. A normally
incident GB is routed by the metasurface before it is launched
normally. Beam axes are indicated by gray dashed lines.
The designed $\bX_s$ elements (b) over an SW-to-PW conversion range
and (c) over a guided SW propagation range.}
 \label{fig:gb_refl_design}
\end{figure}

As a first design example, a translator-reflector for
receiving and launching a Gaussian beam (GB)
propagating normal to the metasurface is presented.
The specific design function is illustrated in
Fig.~\ref{fig:gb_refl_design}(a). A 2-D GB propagating in the
$-y$-axis direction illuminates the metasurface with the beam
axis aimed at $x=-10\lambda$, where $\lambda$ is the free-space
wavelength. The output beam is a GB of the same beam
specification that is launched normally into free space 
from $x=10\lambda$, after being 
translated in the $+x$-axis direction.
The tangent vector of the
input electric field on the surface is specified by
\begin{equation}
\bE^\t{i}_t(x)=\bE^\t{i}_0e^{-(x+10\lambda)^2/2\sigma^2};\ \ 
\bE^\t{i}_0=\zhat~\text{V/m},\ \sigma=2\lambda.
\label{def:Eit_gb_refl}
\end{equation}
Equation~(\ref{def:Eit_gb_refl}) completely specifies the
incident fields in $y\geq 0$.
The minimum waist of the input
beam is equal to $\sqrt{2}\sigma=2\sqrt{2}\lambda$ and it
is located in the $xz$-plane. The tangential vector of the
output beam is specified to be
\begin{equation}
\bE^\t{o}_t(x)=\bE^\t{o}_0e^{-(x-10\lambda)^2/2\sigma^2},\ \ 
\bE^\t{o}_0=\zhat~\text{V/m}.
\label{def:Eot_gb_refl}
\end{equation}
The same amplitude vector and beam waist guarantee that the
output power per unit length in $z$
is equal to that of the input beam.

For each beam, we treat the decayed fields 
at positions away from the
beam axis by more than $3\sigma$, where the field amplitude
reduces to 1.1\% of the peak value, as zero and thus choose
$x_\t{il}=-16\lambda$, $x_\t{iu}=-4\lambda$, $x_\t{ol}=4\lambda$,
and $x_\t{ou}=16\lambda$. Next, the center wavenumber is chosen
to be $k_\t{c}=2k$. Due to the symmetry of the problem, an even
symmetry for $A(x)$ is enforced with respect to $x$. A total
of 17 independent control points, $A_n$ $(n=-16,-15,\ldots,-1)$
in $x\in[x_\t{il},x_\t{iu}]$ and $A_0$ at $x=0$, are assigned.
An equal spacing was chosen between the 16 points,
$x_n$ $(n=-16,-15,\ldots,-1)$. Numerical optimization was
performed for the 17 control point values. In fact, the
power profiles and envelope function shown in
Fig.~\ref{fig:StetmAenv} display the optimization results of
this design drawn to scale. As can be observed in
Fig.~\ref{fig:StetmAenv}(a), $S^\TE_y$ and $S^\TM_y$ 
effectively cancel each other. For the optimized envelope,
it was found that $A_0=16.5~\text{mA/m}$ and other values
scale according to Fig.~\ref{fig:StetmAenv}(b).

The elements of $\bX_s$ of the optimized design were computed
over $-20\lambda<x<20\lambda$. Figure~\ref{fig:gb_refl_design}(b)
shows its four elements in $9\lambda<x<11\lambda$, where
SW-to-PW conversion occurs. They are highly spatially dispersive
functions. The values diverge four times over one wavelength,
where the denominator in (\ref{xs:from_fields}) becomes zero.
The spatial frequency of diverging reactance elements
is a result of the choice $k_\t{c}$ combined with the uniform
phase of $H_{tx}$ associated with broadside launching.
As expected for a lossless design, it is observed that 
$X_{xz}=X_{zx}$. In comparison, the $x$-range near the origin
corresponds to guidance of a TM wave in the absence of any TE
components. The reactance elements over $-\lambda<x<\lambda$ are
plotted in Fig.~\ref{fig:gb_refl_design}(c). Here, with only a
TM-mode SW present, it is noted that $H_{tx}=0$, $E_{tz}=0$,
and $\Re\{E_{tx}H_{tz}^*\}=0$ in (\ref{xs:from_fields}).
A tensor impedance representation is not appropriate in such
a case. In numerical evaluation, (\ref{xs:from_fields}) may be
understood as a limiting case where TE-polarized components
approach zero. Still, some numerical artifact is observed in that
$X_{xz}\neq X_{zx}$ in Fig.~\ref{fig:gb_refl_design}(c). Since
there are no TE-polarized components in this range, only the terms
$X_{xx}$ and $X_{zx}$ are significant. Since $X_{zx}=0$,
no $z$-directed (TE-polarized) electric field will be generated
from an $x$-directed electric current. A constant 
inductive self-reactance at $X_{xx}=1.73\eta$ is consistent
with a TM-mode surface wave having a propagation constant of
$k_\t{c}=2k$. In the guided TM-polarized SW range, the tensor
surface impedance may well be replaced by an isotropic
reactive impedance 
$\bZ_s=\mathbf{I}_tjX_{xx}$
$(\mathbf{I}_t=\text{an identity tensor
in the $xy$-plane})$.
This isotropic choice of $\bZ_s$ over the
constant-amplitude SW range, $x_\t{iu}<x<x_\t{ol}$,
was also tested in numerical analysis. 
The resulting field distributions (not shown)
are found to be the same as the case where the fully-anisotropic
impedance was used.

\begin{figure}[t]
 \centering
 \includegraphics*{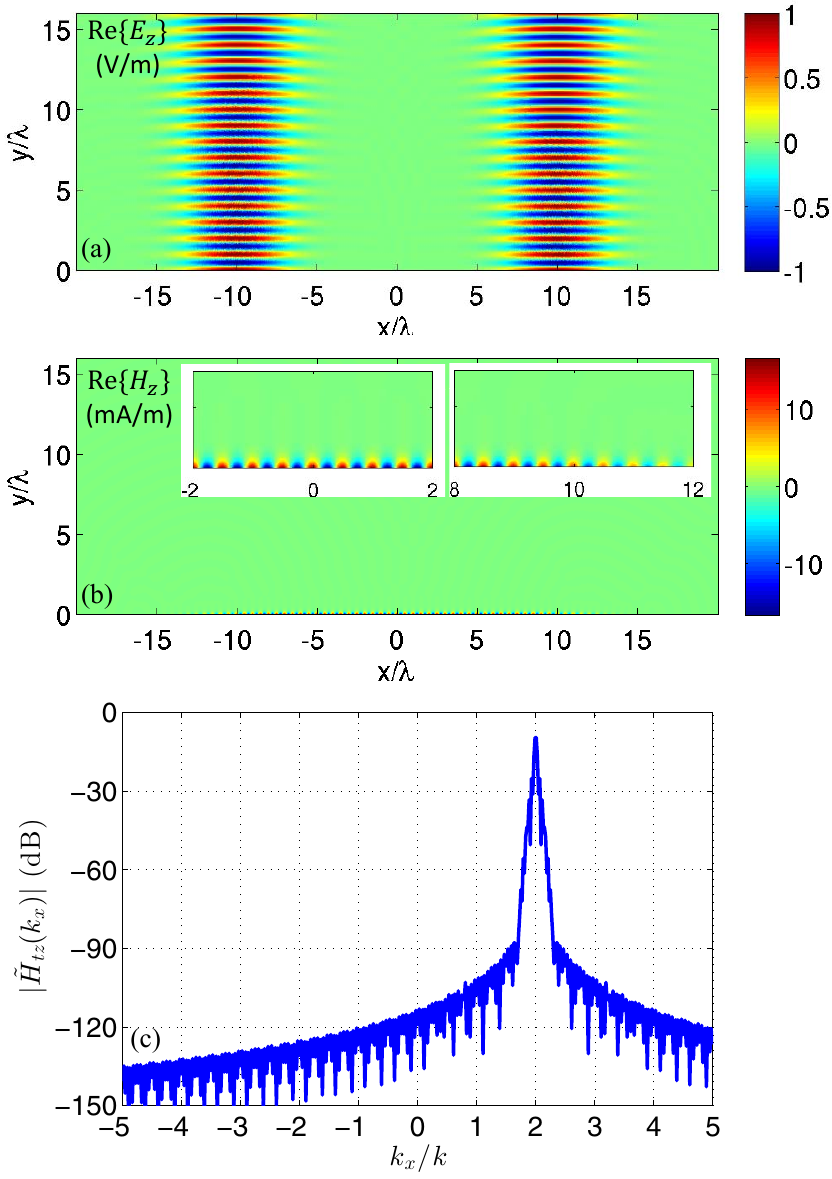}
 \caption{Simulated performance of the GB translator-reflector
metasurface.
(a) A snapshot of $E_z(x,y)$. (b) A snapshot of $H_z(x,y)$.
(c) $|\tilde H_{tz}(k_x)|$.}
 \label{fig:gb_refl_result}
\end{figure}

The scattering characteristics of the metasurface is simulated
using COMSOL Multiphysics. Illuminated by the GB the
metasurface was designed for, Fig.~\ref{fig:gb_refl_result}(a)
plots a snapshot of $E_z$ over an area of $|x|<20\lambda$,
$0<y<16\lambda$. Over the illuminated range by the input GB,
no reflection is seen with only the incident beam field
being observed. In the output GB launch range, the desired
GB with the same strength and beam waist is radiated.
A snapshot of $H_{z}$ is shown in Fig.~\ref{fig:gb_refl_result}(b).
Non-zero fields are observed tightly bound to the $y=0$ surface.
In the inset on the right, a magnified view of the $H_z$ snapshot
is shown over $8\lambda<x<12\lambda$ in
the SW-to-PW conversion range. A surface wave with a diminishing
amplitude is induced as designed. Over the guided SW range
shown in the inset on the left, a TM-polarized SW of a constant
amplitude is clearly visible.
Over the input beam reception range (a magnified view not shown),
the amplitude of the SW gradually grows.
In Fig.~\ref{fig:gb_refl_result}(c), the magnitude of the
SW field spectrum is plotted with respect to $k_x$ in both the
visible $(|k_x|\leq k)$ and invisible $(|k_x|>k)$ ranges.
The spectrum is strongly concentrated around the maximum
at $k_x=k_\t{c}=2k$. The shape of the SW spectrum is completely
determined by the envelope function because
$\tilde H_{tz}(k_x)=\mathscr{F}\{A(x)e^{-jk_\t{c}x}\}=\tilde A(k_x-k_\t{c})$.
Computed using (\ref{PTMy}), the power associated
with the TM-polarized PW component from the visible range
is six orders of magnitude below that of the input GB power.
Hence, the synthesized TM-polarized SW can be considered
completely bound to the reflecting surface.

\subsection{Probe current-fed Gaussian beam launcher}
\label{subsec:launcher}

\begin{figure}[t]
 \centering
 \includegraphics*{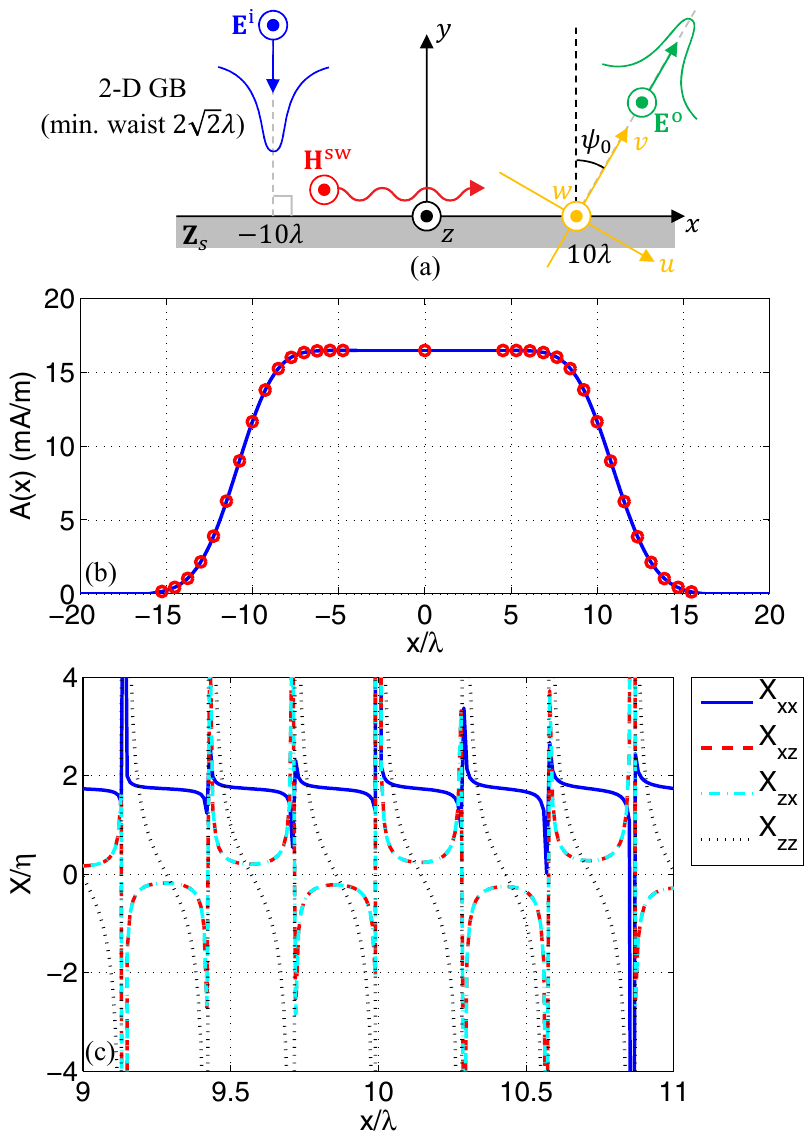}
 \caption{Design of an oblique GB launcher.
(a) An underlying configuration of a beam launcher fed by an
input beam.
(b) The envelop function $A(x)$ and the control points
for $\psi_0=30\degree$.
(c) The reactance tensor elements in the launch range.}
 \label{fig:gb_launch_design}
\end{figure}

A uniform-amplitude SW can be generated by a
localized source. Via reciprocity, an SW can be absorbed by a
localized lossy material or load impedance. When combined with
the PW-to-SW conversion, such a device represents a leaky-wave
antenna that transmits or receives a beam wave.

Here, a GB launcher in an oblique angle is designed.
For this purpose, the GB translator-reflector 
of Sec.~\ref{subsec:reflector} can be adapted to design
an oblique GB launcher excited by an input beam, before the
SW feeding mechanism is replaced with a localized current
probe. The underlying GB launcher configuration is illustrated
in Fig.~\ref{fig:gb_launch_design}(a). The input beam
remains unchanged from the translator-reflector design in
Sec.~\ref{subsec:reflector}. The output beam shape remains the
same, but it is launched at an angle of $\psi_0$ from 
the surface normal from $x=10\lambda$. The output beam can
be first characterized in a rotated $(u,v,w)$ coordinate system
with respect to $z=w$ such that the $v$-axis is aligned
with the output beam axis. Along the $u$-axis, the output
GB is completely characterized by the $w$-component of the
electric field
\begin{equation}
\bE^\t{o}_t(u)=\bE^\t{o}_0e^{-u^2/2\sigma^2},\ \ 
\bE^\t{o}_0=\what~\text{V/m}.
\label{def:Eot_laucher}
\end{equation}
Then, the output beam fields in the entire $uv$-plane can be
obtained by applying the relations in Appendix~\ref{appendix:te}
in the $uv$-plane. A special care should be exercised so that
the fields in the range $v<0$ represent a propagating wave
with $k_v>0$. This can be done by taking a complex conjugate
of the fields from Appendix~\ref{appendix:te}, corresponding
to a time-reversal transformation. Fields evaluated along
the $x$-axis in the $uv$-plane can be transformed to the 
$xyz$ system by a proper coordinate rotation, 
before they can be used for the device design in the $xy$-plane.

For a 30\textdegree{} GB launcher $(\psi_0=30\degree)$, the
$x$-range of numerical synthesis of the envelope function was set
with $x_\t{il}=-16\lambda$, $x_\t{iu}=-4\lambda$,
$x_\t{ol}=3.75\lambda$, and $x_\t{ou}=16.25\lambda$. For the
output beam, the range was slightly extended to account for the
oblique launch angle. In each of $x\in[x_\t{il},x_\t{iu}]$ and
$x\in[x_\t{ol},x_\t{ou}]$, 15 equally-spaced
control points were assigned. Hence, together with $A_0$ assigned
to $x=0$, a total of 31 control points were defined for $A(x)$
and they were optimized for the minimum square error (\ref{def:sqerr}).
The center wavenumber was chosen to be $k_\t{c}=2k$.
The optimized envelope function together with the control points
are plotted in Fig.~\ref{fig:gb_launch_design}(b).
The amplitude of the guided SW was found to be
$A_0=16.5~\text{mA/m}$, which remains unchanged from the
translator-reflector design in Sec.~\ref{subsec:reflector}
as can be expected. 
In Fig.~\ref{fig:gb_launch_design}(c), the reactance
tensor elements over $9\lambda<x<11\lambda$ in the oblique GB
launch range are plotted. They are different from
Fig.~\ref{fig:gb_refl_design}(b) and they diverge less frequently
with respect to $x$. This is due to different phase gradients
present in $H_{tx}$ in (\ref{xs:from_fields}) for the same
SW expected in the two designs. Indicative of a lossless property,
$X_{xz}=X_{zx}$ is observed.

\begin{figure}[t]
 \centering
 \includegraphics*{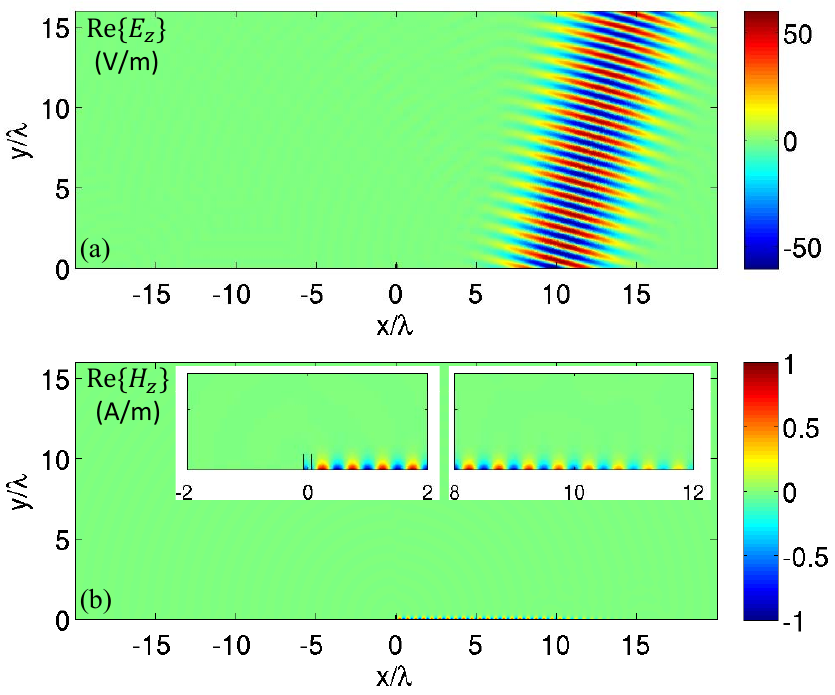}
 \caption{Simulation results for the oblique GB launcher.
(a) A snapshot of $E_z(x,y)$.
(b) A snapshot of $H_z(x,y)$.}
 \label{fig:gb_launch_result}
\end{figure}

Since launching the output GB using a localized source is
of interest, we adopt a localized source in the form of a two-element
array around $x=0$.
{\color{red}In other words, the $+x$-propagating constant-amplitude
SW near $x=0$ is provided by a current probe instead of an
input GB.} 
Two elementary radiators separated by a
quarter wavelength with a 90\textdegree{} phase difference can
create a unidirectional radiation pattern~\cite{balanis2005}.
Hence, denoting the attenuation constant of the
SW around $x=0$ by $\alpha_\t{c}$,
we use an impressed electric current source given by
\begin{equation}
\bJ_\t{imp}=\yhat J_0e^{-\alpha_\t{c}y}
\left[\delta\left(x+\frac{\lambda_\t{c}}{8}\right)e^{j\frac{\pi}{4}}
+\delta\left(x-\frac{\lambda_\t{c}}{8}\right)e^{-j\frac{\pi}{4}}\right]
\label{def:Jimp}
\end{equation}
in $0\leq y\leq\lambda_\t{c}/4$ and zero in $y>\lambda_\t{c}/4$, 
where $\lambda_\t{c}=2\pi/k_\t{c}$ is the guided wavelength of the SW.
This model represents two strips of vertical electric current
of a $\lambda_\t{c}/4$ height. Their $y$-dependence is matched to
that of the TM SW for maximum coupling. In simulation results
that follow, $J_0=1~\text{A/m}$ has been used.

Snapshots of the $z$-component of the total electric and
magnetic fields are plotted in Fig.~\ref{fig:gb_launch_result}.
In Fig.~\ref{fig:gb_launch_result}(a), a GB is launched
from $x=10\lambda$ at an angle of 30\textdegree{} as desired.
In the snapshot of $H_z$ in Fig.~\ref{fig:gb_launch_result}(b),
the TM-mode wave is visible only as an evanescent wave bound
to the surface. The left inset shows a magnified view
near the two-element current source at $x=0$. The two short line
segments in black indicate the source positions and their height.
A constructive
interference between the fields generated by the two strip currents
produces a $+x$-propagating SW with a maximum amplitude of
1~A/m (realized on the surface)
as designed. A destructive interference results in
negligible excitation of an SW toward the $-x$-axis direction.
The inset on the right shows a portion of the SW-to-PW conversion
range. The SW amplitude is gradually reduced with increasing $x$
due to continuous leakage into space.

\subsection{Gaussian beam-excited focusing lens}
\label{subsec:lens}

\begin{figure}[t]
 \centering
 \includegraphics*{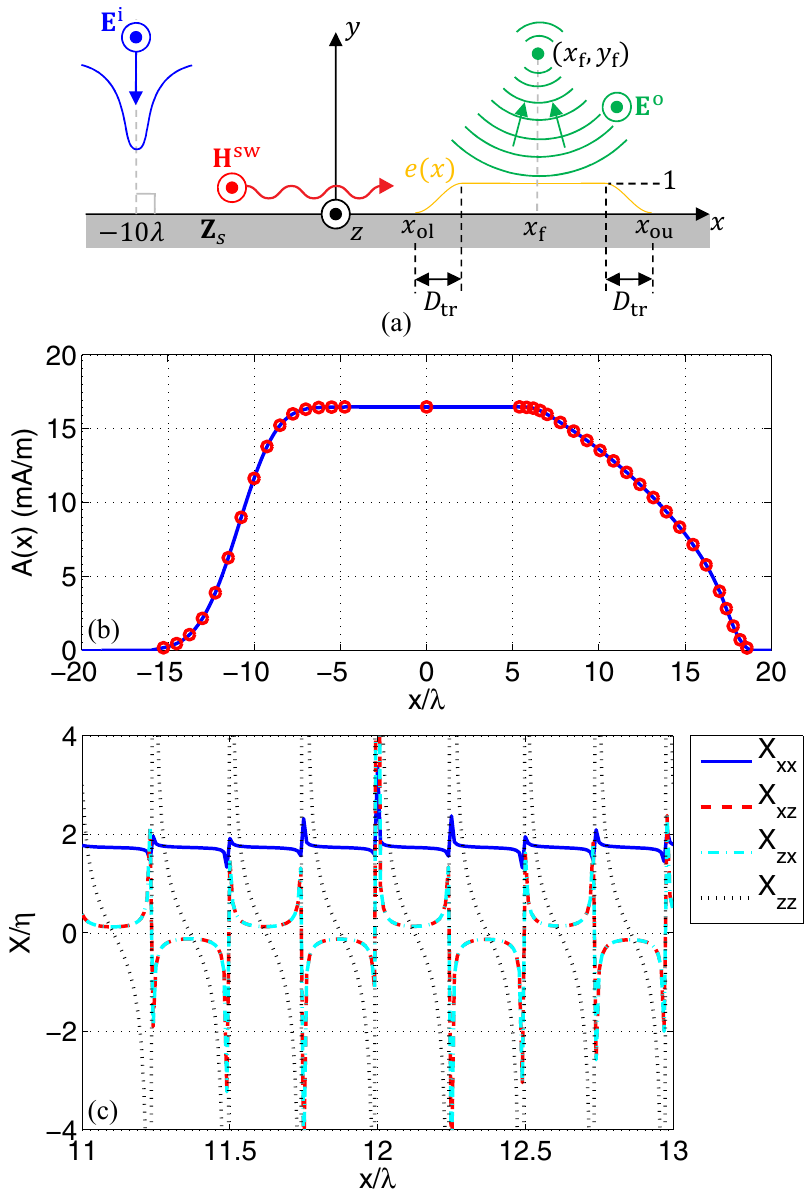}
 \caption{A flat lens excited by an input GB.
(a) The GB-to-SW-to-converging wave transformation configuration.
(b) The envelope function $A(x)$ and the 38 control points
of the optimized design.
(c) The reactance tensor elements in the cylindrical wave
launch range.} 
 \label{fig:gb_lens_design}
\end{figure}

In order to demonstrate the versatility of the design technique,
a focusing lens excited by the same normally incident GB considered
in Sec.~\ref{subsec:reflector} is designed. The desired
wave transformation is illustrated in Fig.~\ref{fig:gb_lens_design}(a).
A normally incident GB is converted into a TM-polarized SW
in $x<0$. The converted SW is transformed into a converging
cylindrical wave in the range $x_\t{ol}<x<x_\t{ou}$, where an
envelop function $e(x)$ for $|E_{tz}(x)|$ is introduced.
On both ends within the range, a transition period of a length 
$D_\t{tr}$ in a raised cosine profile is defined for a smooth
transition between unity and zero, as illustrated
in Fig.~\ref{fig:gb_lens_design}(a). Specifically,
\begin{equation}
e(x)=\begin{cases}
0, & x<x_\t{ol}\ \  \text{or}\ \  x\geq x_\t{ou}\\
[1-\cos(\frac{x-x_\t{ol}}{D_\t{tr}/\pi})]/2,
& x_\t{ol}\leq x<x_\t{ol}+D_\t{tr}\\
1, & x_\t{ol}+D_\t{tr}\leq x<x_\t{ou}-D_\t{tr}\\
[1-\cos(\frac{x-x_\t{ou}}{D_\t{tr}/\pi})]/2, 
& x_\t{ou}-D_\t{tr}\leq x<x_\t{ou}
\end{cases}.
\label{def:eo}
\end{equation}
In order for the launched cylindrical wave to converge at a focal
point $(x_\t{f},y_\t{f})$, a phase distribution for $E_{tz}$
is introduced as
\begin{equation}
\phi(x)=k\sqrt{(x-x_\t{f})^2+y_\t{f}^2}+\phi_0,
\label{def:phito}
\end{equation}
where $\phi_0$ is an arbitrary phase constant. Then, the tangential
electric field for the output wave is defined to be
\begin{equation}
\bE^\t{o}_t(x)=\bE^\t{o}_0e(x)e^{j\phi(x)},\ \ 
\bE^\t{o}_0=\zhat E^\t{o}_0.
\label{def:Eot_lens}
\end{equation}
The amplitude $E^\t{o}_0$ is determined to make the overall
conversion from the input GB to the output focusing wave lossless.

For the TE-polarized GB input wave considered in 
Sec.~\ref{subsec:reflector}, a focusing lens with a focal point at
$(x_\t{f},y_\t{f})=(12\lambda,10\lambda)$ is designed using an
SW-to-PW conversion range with $x_\t{ol}=5\lambda$,
$x_\t{ou}=19\lambda$. With a choice of $D_\t{tr}=2\lambda$,
the $x$-range of a constant-amplitude electric field for the
output wave is $7\lambda<x<17\lambda$. The phase constant $\phi_0$
is set to zero. For this chosen profile of $\bE^\t{o}_t(x)$, by
setting the output PW power equal to the input GB power, the
amplitude is found to be $E^\t{o}_0=0.569~\text{V/m}$.
Next, $N_\t{ci}=15$ control points are defined for the input range
$-16\lambda<x<-4\lambda$, four control points are assigned to
both transition ranges in the output range. To the constant-amplitude
range of $10\lambda$ length, 14 control points are assigned.
The total number of control points is 38 for this lens design.
The envelop function $A(x)$ of the optimized design is
shown in Fig.~\ref{fig:gb_lens_design}(b) together with the
control points indicated by red circles.
In $7\lambda<x<17\lambda$, where a TE-polarized aperture
electric field of a uniform magnitude is synthesized,
it is observed that
$A(x)$ diminishes with an increasingly negative slope with
respect to $x$ as the SW power is continuously lost to the output
PW. The four reactance element values of the optimized design
over $11\lambda<x<13\lambda$ in the SW-to-PW conversion range 
are shown in Fig.~\ref{fig:gb_lens_design}(c) as highly
spatially dispersive functions. The lossless nature of the
metasurface lens is indicated by the fact $X_{xz}=X_{zx}$.

\begin{figure}[t]
 \centering
 \includegraphics*{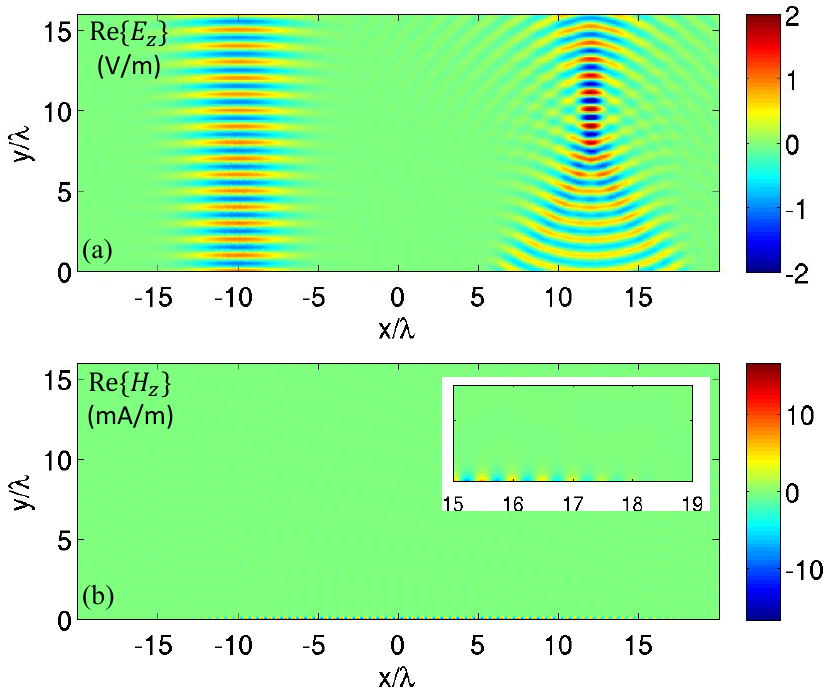}
 \caption{Simulation results for the flat focusing lens
excited by GB illumination.
(a) A snapshot of $E_z(x,y)$.
(b) A snapshot of $H_z(x,y)$.}
 \label{fig:gb_lens_result}
\end{figure}

Figure~\ref{fig:gb_lens_result} shows the simulated results for
the lens obtained using COMSOL Multiphysics. A snapshot of the
TE-component of the electric field is plotted in
Fig.~\ref{fig:gb_lens_result}(a). Converging cylindrical
waves form a focal point at 
$(x,y)=(x_\t{f},y_\t{f})=(12\lambda,10\lambda)$ as designed.
A snapshot of $H_z(x,y)$ in Fig.~\ref{fig:gb_lens_result}(b)
shows the TM-polarized field, which is tightly bound to the metasurface.
In a magnified view in the range $15\lambda<x<19\lambda$,
the SW wave is seen to be diminishing and disappearing
as all of its guided power along the $+x$-direction is leaked
into the PW spectrum.

\section{Conclusion}
\label{sec:conclusion}

Arbitrary polarization-preserving beam control using
passive, lossless
impenetrable metasurfaces has been presented. It is based on
a custom SW in an orthogonal polarization designed
to propagate along the surface. By finding an optimal spatial
envelope for the SW such that there is zero normal component for
the Poynting vector everywhere on the surface,
a lossless metasurface can convert a PW into an SW,
guide the SW unattenuated along the surface, and launch a
PW via an SW-to-PW transformation. As specific design examples,
a GB translator-reflector, a GB launcher excited by a localized source,
and a GB-fed flat focusing lens have been presented.
Lossless metasurfaces characterized by a reactance tensor
in a plane
may be realized using an array of rotated subwavelength resonators
printed on a grounded dielectric substrate.

Availability of an orthogonal polarization for SW synthesis
simplifies the design process owing to the orthogonality of
power flow in the surface normal direction between
the PW and SW on the metasurface. The same design approach
can be envisioned for polarization-converting transformations
between the input and output beams or for using an SW that is
co-polarized with the input and output beams. SW synthesis
in such a scenario
is expected to be challenging due to the complex interference
power pattern created by the co-polarized
PW and SW components in the system.

{\color{red}At microwave frequencies, a planar array of
rotated anisotropic printed conductor patch resonators 
on a grounded dielectric substrate is a promising
configuration~\cite{fong_ieeejap2010,minatti_ieeejap2012,patel_ieeejap2013,
lee_ieeejap2016} that allows accurate, low-cost fabrication
for realizing the inhomogeneous, tensor surface impedances
in this study.} 

\appendix
\section{Spectral representations of fields and power density
in TE polarization}
\label{appendix:te}

For TE-polarized fields, the tangential electric field on the
surface, $E_{tz}(x)=E_z(x,y=0)$, is expressed as an inverse Fourier
transform of its spectrum $\tilde E_{tz}(k_x)$ as
\begin{equation}
E_{tz}(x)=\mathscr{F}^{-1}[\tilde E_{tz}(k_x)]
=\frac{1}{2\pi}\int\limits_{-\infty}^\infty\tilde E_{tz}(k_x)e^{-jk_xx}dk_x,
\label{Etz:invF}
\end{equation}
where $k_x$ is the wavenumber in the $x$-direction.
{\color{red}Here, we add a tilde to a spatial function to indicate its
Fourier transform in $x$.}
Equation~(\ref{Etz:invF}) represents a superposition 
of plane waves propagating
in the $xy$-plane over the $x$-wavenumber $k_x$, evaluated at $y=0$.
Hence, at any point $(x,y)$ in $y\geq 0$, the field $E_z(x,y)$
is the same superposition of plane waves evaluated at $(x,y)$ after
a proper phase delay in the $y$ direction is incorporated to
each plane wave component
in the integrand of (\ref{Etz:invF}). In other words,
\begin{equation}
E_z(x,y)=\frac{1}{2\pi}\int\limits_{-\infty}^\infty
\tilde E_{tz}(k_x)e^{-j(k_xx+k_yy)}dk_x,
\label{Ez:invF}
\end{equation}
where
\begin{equation}
k_y=\begin{cases}
\sqrt{k^2-k_x^2}, & |k_x|\leq k\\
-j\sqrt{k_x^2-k^2}, & |k_x|>k
\end{cases}
\label{def:ky}
\end{equation}
and $k$ is the free-space wavenumber. 
At an observation point $(x,y)$, (\ref{Ez:invF}) represents
a superposition of propagating and evanescent plane waves.
{\color{red}In (\ref{def:ky}), a choice of the negative branch
for the square root for $|k_x|>k$
guarantees an exponential decay toward
$y\to\infty$ for each inhomogeneous plane wave component.}
The associated magnetic field components in $y\geq 0$ are
obtained from Maxwell's curl equations as
\begin{eqnarray}
H_x(x,y) &=& \frac{1}{2\pi}\int\limits_{-\infty}^\infty\frac{k_y}{k\eta}
\tilde E_{tz}(k_x)e^{-j(k_xx+k_yy)}dk_x,
\label{Hx:invF}\\
H_y(x,y) &=& -\frac{1}{2\pi}\int\limits_{-\infty}^\infty\frac{k_x}{k\eta}
\tilde E_{tz}(k_x)e^{-j(k_xx+k_yy)}dk_x,
\label{Hy:invF}
\end{eqnarray}
where $\eta$ is the free-space intrinsic impedance.

The normal component of the time-average Poynting vector is the
key quantity in the design process.
Using $H_{tx}(x)=H_x(x,y=0)$ from (\ref{Hx:invF}),
the $y$-component of the TE-mode Poynting vector
is expressed as
\begin{equation}
\begin{split}
S^\TE_y(x) &= \frac{1}{2}\Re\{E_{tz}H_{tx}^*\}
=\frac{1}{2}\Re\{E_{tz}^*H_{tx}\}\\
&= \frac{1}{2}\Re\left\{E_{tz}^*(x)\frac{1}{2\pi}
\int\limits_{-\infty}^\infty\frac{k_y}{k\eta}
\tilde E_{tz}(k_x)e^{-jk_xx}dk_x\right\}.
%&= \frac{1}{2k\eta}\Re\left\{E_{tz}^*(x)
%\mathscr{F}^{-1}\left[k_y\tilde E_{tz}(k_x)\right]\right\}.
\end{split}
\label{STEy}
\end{equation}
For a given profile $E_{tz}(x)$, evaluation of (\ref{STEy}) involves
two one-dimensional integrals, of which numerical evaluation
can be performed efficiently.

\section{Spectral representations of fields and power density
in TM polarization}
\label{appendix:tm}

Expressions of field components and the normal Poynting vector
component in the spectral domain in terms of 
$H_{tz}(x)=H_z(x,y=0)$ are
derived following the same procedure as in Appendix~\ref{appendix:te}.
Here, only the final results are summarized.

In $y\geq 0$, the three TM-mode field components,
$H_z$, $E_x$, and $E_y$, are expressed as a superposition 
of plane waves as
\begin{eqnarray}
H_z(x,y) &=& \frac{1}{2\pi}\int\limits_{-\infty}^\infty
\tilde H_{tz}(k_x)e^{-j(k_xx+k_yy)}dk_x,
\label{Hz:invF}\\
E_x(x,y) &=& -\frac{1}{2\pi}\int\limits_{-\infty}^\infty\frac{\eta k_y}{k}
\tilde H_{tz}(k_x)e^{-j(k_xx+k_yy)}dk_x,
\label{Ex:invF}\\
E_y(x,y) &=& \frac{1}{2\pi}\int\limits_{-\infty}^\infty\frac{\eta k_x}{k}
\tilde H_{tz}(k_x)e^{-j(k_xx+k_yy)}dk_x,
\label{Ey:invF}
\end{eqnarray}
where a tilde notation has been used to indicate the Fourier transform
of $H_{tz}(x)$. The expression for the $y$-component of the TM-mode
Poynting vector is found to be
\begin{equation}
S^\TM_y(x)
=\frac{1}{2}\Re\left\{H_{tz}^*(x)\frac{1}{2\pi}
\int\limits_{-\infty}^\infty\frac{\eta k_y}{k}
\tilde H_{tz}(k_x)e^{-jk_xx}dk_x\right\}.
\label{STMy}
\end{equation}

%\bibliographystyle{apsrev4-1}
%\nocite{apsrev41Control}
%\bibliography{IEEEabrv,references}


\end{document}